\documentclass[10pt,aps,prl,twocolumn,showpacs,amsmath,amssymb]{revtex4-2}
\usepackage{graphicx} 
\usepackage{xcolor}
\usepackage{amsmath}
 \usepackage{hyperref}
\usepackage[top = 2cm, bottom = 2cm, right = 2cm, left =2cm]{geometry}

\begin{document}

\title{A New Cloud of Strings}

\author{G. Alencar} 
\email{ geova@fisica.ufc.br}
\affiliation{Departamento de F\'isica, Universidade Federal do Cear\'a, Caixa Postal 6030, Campus do Pici, 60455-760 Fortaleza, Cear\'a, Brazil}

\author{R. R. Landim}
\email{renan@fisica.ufc.br} 
\affiliation{Departamento de F\'isica, Universidade Federal do Cear\'a, Caixa Postal 6030, Campus do Pici, 60455-760 Fortaleza, Cear\'a, Brazil}

 \author{R. N. Costa Filho}
\email{rai@fisica.ufc.br}
\affiliation{Departamento de F\'isica, Universidade Federal do Cear\'a, Caixa Postal 6030, Campus do Pici, 60455-760 Fortaleza, Cear\'a, Brazil}

\date{\today}

\begin{abstract}
In this work, we present a generalization of the cloud of strings model originally proposed by Letelier~\cite{Letelier:1979ej}, by introducing a \textit{magnetic-like} component, \( \Sigma_{23} \), in addition to the \textit{electric-like} component, \( \Sigma_{01} \), considered in the original formulation. This extension leads to a new black hole solution of the form
\begin{equation}
    f(r) = 1 - \frac{2M}{r} + \frac{g_s^2 \ell_s^2}{r^2} \, {}_2F_1\left(-\frac{1}{2}, -\frac{1}{4}, \frac{3}{4}, -\frac{r^4}{\ell_s^4}\right),
\end{equation}
where \( {}_2F_1 \) denotes the Gaussian hypergeometric function. The solution is characterized by the string length \( \ell_s \) and the string coupling constant \( g_s \). Within this framework, the original Letelier solution is recovered in the point-particle limit \( \ell_s \to 0 \), with the parameter \( a \) naturally identified as \( a = g_s^2 \). This identification yields the condition \( 0 < a < 1 \), which now emerges from the model itself, ensuring both energy positivity and the existence of horizons—rather than being imposed ad hoc. We also investigate the thermodynamic properties of the resulting black hole, and find that the entropy remains consistent with the Bekenstein–Hawking formula, \( S = A/4 \), in agreement with several string-theoretic derivations.
\end{abstract}

\maketitle

\section{Introduction}

Recent years have witnessed significant advances in General Relativity, driven by groundbreaking observations such as gravitational waves from binary mergers involving black holes~\cite{LIGOScientific:2016aoc} and black hole-neutron star systems~\cite{LIGOScientific:2017vwq}, as well as direct imaging of supermassive black hole environments in M87~\cite{EventHorizonTelescope:2019dse} and the Milky Way~\cite{EventHorizonTelescope:2022wkp}. These achievements mark the dawn of the multimessenger era, offering unprecedented opportunities to probe compact objects and test the strong-field regime of gravity~\cite{Addazi:2021xuf}.

In another direction, black holes interacting with surrounding matter provide critical insights into the interplay between geometry and matter. Among these, the cloud of strings model, introduced by Letelier~\cite{Letelier:1979ej}, offers an elegant framework for describing spacetimes influenced by a continuous distribution of strings. This model has been extensively applied to explore the impact of extended matter sources on black hole spacetimes~\cite{Ganguly:2014cqa, Panotopoulos:2018law, Singh:2020nwo, Mustafa:2022xod, Ghosh:2014dqa,deMToledo:2018tjq,Toledo:2018hav}, highlighting its importance as a tool for understanding such interactions.

In this work, we generalize the cloud of strings model by introducing an additional \textit{magnetic-like component}, $\Sigma_{23}$. This modification results in a complete stress-energy tensor of the form $\mathrm{diag}(-\rho, -\rho, p, p)$, parameterized by two constants and governed by a unique equation of state. We investigate the resulting black hole solutions, examining how these changes influence the spacetime geometry and thermodynamics. This provides a new perspective on the interaction between extended matter distributions and gravitational fields.

The starting point is the Nambu–Goto action, expressed as:
\begin{equation}
    S_{NG}=\int \sqrt{-\gamma}\mathcal{M}d\lambda^0d\lambda^1,\;\gamma_{AB}=g_{\mu\nu}\frac{\partial x^\mu}{\partial \lambda^A}\frac{\partial x^\nu}{\partial \lambda^B}\ .\label{ActionCS1}
\end{equation}
Here, $\lambda^0$ and $\lambda^1$ are timelike and spacelike parameters, respectively; $\mathcal{M}$ is a dimensionless constant that characterizes the string; and $\gamma$ is the determinant of the induced metric $\gamma_{AB}$ on the submanifold. Using the following definitions:
\begin{equation}\label{Sigma}
 \Sigma^{\mu\nu}=\epsilon^{AB}\frac{\partial x^\mu}{\partial \lambda^A}\frac{\partial x^\nu}{\partial \lambda^B},\;  \gamma= \frac{1}{2}\Sigma^{\mu\nu}\Sigma_{\mu\nu}\ ,
\end{equation}
the stress-energy tensor is written as:
\begin{equation}\label{STCS}
 T^{\mu}_{\nu}=\rho_p\frac{\Sigma^{\ \mu\alpha}\Sigma_{\alpha\nu}}{8\pi\sqrt{-\gamma}}\ ,  
\end{equation}
where $\rho_p$ is the proper density and $\Sigma^{\mu\nu}$ satisfies the following equations:
\begin{equation}
    \nabla_\mu \left(\rho_p \Sigma^{\mu\nu}\right)=0,\;\Sigma^{\mu\beta}\nabla_\mu\left[\frac{{\Sigma_{\beta}}^{\nu}}{(-\gamma)^{1/2}}\right]=0\ .\label{conserv1}
\end{equation}

For spherically symmetric solutions, Letelier showed that the only component satisfying $\gamma<0$ is the "electric-like" component, $\Sigma_{01}$. The corresponding black hole solution is:
\begin{equation}\nonumber
    f(r)=1-|a|-\frac{2M}{r}
\end{equation}
Depending on the value of $a$, this solution may feature one or no horizons. Moreover, it lacks a remnant mass or entropy. However, by introducing the "magnetic-like" component $\Sigma_{23}$, the condition $\gamma<0$ is preserved, and a more general stress-energy tensor is obtained while maintaining the desired physical properties.

\section{Stress-Tensor and Energy Conditions}
To find the complete solution, we consider a general spherically symmetric metric of the form:
\begin{equation}\label{general}
    ds^2=-A(r)dt^2+B(r)dr^2+C^2(r)\left(d\theta^2+\sin^2\theta d\phi^2\right)\ .
\end{equation}
In this scenario, Eq.~\eqref{conserv1} leads to:
\begin{equation}\label{conserv2}
    \partial_\mu\left(C^2\sqrt{AB}\sin\theta\rho \Sigma^{\mu\nu}\right)=0\ .
\end{equation}
The general solutions of Eq.~\eqref{conserv2}, consistent with spherical symmetry, are:
\begin{equation}\label{gensol}
    \rho_p\Sigma_{01}=\frac{a\sqrt{AB}}{C^2},\; \rho_p\Sigma_{23}=b(r)\sin\theta\ ,
\end{equation}
where $a$ is a constant and $b(r)$ is an arbitrary function. The non-zero components of $T^\mu_{\ \nu}$ are:
\begin{align}
    T^{0}_{\ 0}=&T^{1}_{\ 1}=-\frac{a^2}{C^2\sqrt{a^2-b^2}}\ ,\nonumber
    \\
    T^{2}_{\ 2}=&T^{3}_{\ 3}=\frac{b^2}{C^2\sqrt{a^2-b^2}}\ .\label{T}
\end{align}
From the conservation law, $b(r)$ is determined as:
\begin{equation}\label{bsol}
   2C'\left(T^{0}_{\ 0}-T^{2}_{\ 2}\right)+\frac{dT^{0}_{\ 0}}{dr}=0\rightarrow b(r)=\frac{a c_0^2}{\sqrt{c_0^4+C^4}}\ ,
\end{equation}
where $c_0$ is a constant. The unique stress-energy tensor is then:
\begin{align}
    \rho=&-p_r=\frac{|a|}{C^4}\sqrt{c_0^4+C^4}\ ,\label{rhopr}
    \\
    p_l=&T^{2}_{\ 2}=T^{3}_{\ 3}=\frac{|a|c_0^4}{C^4\sqrt{c_0^4+C^4}}\ .\label{pl}
\end{align}
In the limit $c_0=0$, this stress tensor reduces to Letelier’s original model, with $a \to |a|$ \cite{Letelier:1979ej}.  Finally, Eq.~\eqref{bsol} ensures:
\begin{equation}
    \gamma=\frac{-a^2}{\rho^2(c_0^4+C^4)}<0\ ,
\end{equation}
as required, providing a fully consistent model.

The study of the energy conditions is organized below:
\begin{enumerate}
    \item \textbf{Null energy condition}: $\rho+p_i\geq0.$\\
From Eq. (\ref{rhopr}) we see that $\rho+p_r=0$ and, from Eq. (\ref{pl}), we see that $\rho+p_l$ is positive definite. Therefore, the NEC is satisfied.
\item  \textbf{Weak energy condition}: NEC + $\rho\geq 0$.\\
From Eq. (\ref{rhopr}) we see that $\rho\geq 0$. Since the NEC is satisfied, we have the WEC satisfied.
\item \textbf{Strong energy condition}: $\rho+\sum p_i\geq0.$\\
From Eqs. (\ref{rhopr}) and (\ref{pl}), we have that $\rho+\sum p_i=2p_l\geq 0$. Therefore, the SEC is satisfied.
\item \textbf{Dominant energy condition}: $	\rho \geq |p_i|$. \\
From Eq.  (\ref{rhopr}) we have that $\rho = |p_r|$ and 
\begin{equation}
    \frac{\rho}{p_l}=1+\frac{C^4}{c_0^4}\geq1.
\end{equation}
With this, the DEC is also satisfied.
\end{enumerate}
Finally, we conclude that all energy conditions are satisfied.

At this point, it is important to interpret the physical meaning of the constants \( a \) and \( c_0 \).  
From dimensional analysis, we find that \( [a] \sim [L]^0 \) and \( [c_0] \sim [L]^2 \),  
which naturally suggests identifying \( c_0 \) with the string length \( \ell_s \), up to numerical factors.  
In the limit of tensionless strings, a non-vanishing stress-energy tensor remains, as expected,  
which supports this identification. Therefore, from now on, we adopt \( c_0 = \ell_s \). 

The interpretation of \( a \), however, is less direct since it is dimensionless.  
To clarify its role, we note that in the regime \( r \gg \ell_s \),  
the correction to the energy density behaves as \( \sim a \ell_s^4 / r^6 \).  
This is precisely the type of correction that arises in the effective Dirac–Born–Infeld (DBI) action derived from string theory,  where one finds \( a \sim g_s^2 \) \cite{Born:1934gh,Dirac:1962iy,Fradkin:1985qd}.  Moreover, in the limit \( a \to 0 \), the entire stress-energy tensor vanishes, indicating that \( a \) encodes the strength of the string backreaction.  Since we are modeling a cloud of strings, and the gravitational coupling in string theory is given by \( G_N \sim g_s^2 \ell_s^2 \), it is natural to identify \( a \sim g_s^2 \) as the effective coupling controlling the gravitational contribution of the string cloud at tree level.  
This interpretation will be further reinforced below.

\section{Black Holes}
With the above identifications, the Einstein Equations give us
\begin{equation}\label{einsteinb00}
    \frac{d(rf)}{dr}=1+r^2T^0_{\ 0}=1-\frac{g_s^2}{r^2}\sqrt{\ell_s^4+r^4}
\end{equation}
\begin{equation}\label{einsteinb22}
    \frac{d(r^2f')}{dr}=2r^2T^2_{\ 2}=2\frac{g_s^2\ell_s^4}{r^2\sqrt{\ell_s^4+r^4}}
\end{equation}
Multiplying by $r$ and differentiating the equation (\ref{einsteinb00}) with respect to $r$ and using (\ref{einsteinb22}), with $f(r)=1+F(-r^4/c_0^4)/r^2$, we have
\begin{equation}\label{hyperF}
z(1-z)F''(z)+(\frac{3}{4}-\frac{z}{4})F'(z)-\frac{F(z)}{8}=0,
\end{equation}
with $z=-r^4/\ell_s^4$. This is a Hypergeometric differential equation with $a=-1/2, b=-1/4,c=3/4$. We have two solutions around each of the points $ z=0$, $ z=1$, and $z=\infty$. Considering at $z=0$ we  obtain
\begin{align}
    F(z)=&A_0z^{1-c}\ {}_2F_1(a-c+1,b-c+1,2-c,z)\nonumber\\
    &+B_0\ {}_2F_1(a,b,c,z),\label{hyperF2}
\end{align}
were $A_0$ and $B_0$ are complex constants. Substituting the values of $a,b,c$, we see that the first term of the equation (\ref{hyperF2}) reduces to $A_0z^{1/4}=Ar$, where $A$ is a real constant. Making $A=-2M$, we arrive at
\begin{equation}
    f(r)=1-\frac{2M}{r}+B \ {}_2F_1(-1/2,-1/4,3/4,-r^4/\ell_s^4)/r^2,
\end{equation}
In order to obtain the constant $B$, we use the equation  (\ref{einsteinb00}) and take the limit with $r\rightarrow 0$:
$$
g_s^2\ell_s^2=\lim_{r\rightarrow 0}r^2\frac{d(rf)}{dr}=B,
$$
to finally arrive at
\begin{equation}\label{frsol}
    f(r)=1-\frac{2M}{r}+\frac{g_s^2\ell_s^2}{r^2} \ {}_2F_1(-1/2,-1/4,3/4,-\frac{r^4}{\ell_s^4}).
\end{equation}

The horizon is located at the roots of:
\begin{equation}\label{mass}
    2M - r_h = \frac{g_s^2\ell_s^2}{r_h} \ {}_2F_1\left(-\frac{1}{2}, -\frac{1}{4}, \frac{3}{4}, -\frac{r_h^4}{\ell_s^4}\right).
\end{equation}
To better understand, the horizons can be obtained as zero of 
\begin{equation}
    rf(r)=r-2M+\frac{g_s^2\ell_s^2}{r} \ {}_2F_1(-1/2,-1/4,3/4,-r^4/\ell_s^4),
\end{equation}
since $r>0$.
From equation (\ref{einsteinb00}) we have
\begin{equation}
    \frac{d(rf(r))}{dr}=1-\frac{g_s^2}{r^2}\sqrt{\ell_s^4+r^4}.
\end{equation}
We conclude that if $g_s^2\ge1$, $f'(r_h)\le0$, this implies that we have only one horizon. If $g_s^2<1$, $f'(r_h)$ can be positive, zero, or negative, and we have two horizons, $r_{-}$ and $r_{+}$, where
\begin{equation}\label{horiz2}
    r_{-}<r_0<r_{+}, \quad r_0=\frac{\ell_s g_s}{(1-g_s^4)^{1/4}},
\end{equation}
with $f'(r_{-})<0$ and $f'(r_{+})>0$. If $f(r_0)=0$, this implies $f'(r_0)=0$, and we have only one horizon.

While it is not possible to solve for \(r_h\) explicitly, the analysis shows that the solution features one horizon for \(g_s^2 \geq 1\) and two horizons for \(g_s^2 < 1\). The asymptotic behavior of the solution is:
\begin{equation}
    f(r) = 1 - \frac{2M}{r} - g_s^2 + \frac{g_s^2\ell_s^4}{6r^4}, \quad r \gg \ell_s. \label{frsolrbig}
\end{equation}
This demonstrates that the solution is asymptotically flat and reduces to the Letelier solution in the limit \(\ell_s \to 0\).
Even in this limit, obtaining an analytical expression for the event horizon is highly nontrivial, as it ultimately requires solving a quartic polynomial equation.  We show bellow the mass $M$ in function of $r_h$ and the parameters $\ell_s, g_s$. 
\begin{figure}[h]
 \includegraphics[width=8cm]{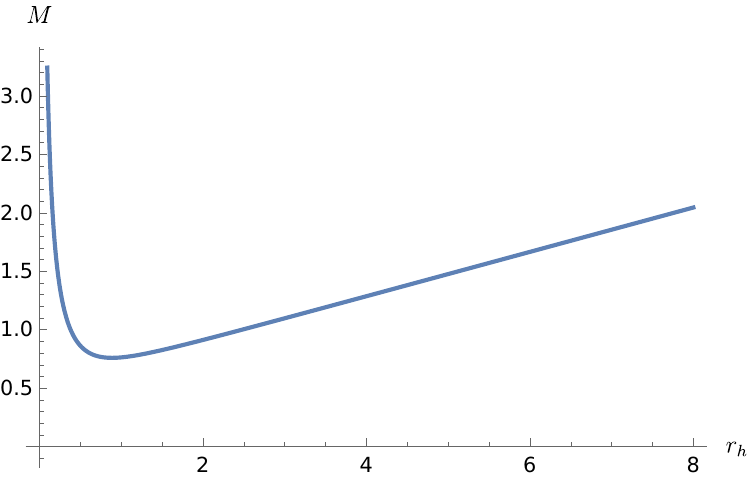}
 \caption{The   mass in function of the $r_h$  with $\ell_s=1$ and $g_s=0.62$.}\label{fig3}
 
\end{figure}

\section{Thermodynamics}
\begin{figure}[h]
\includegraphics[width=8cm]{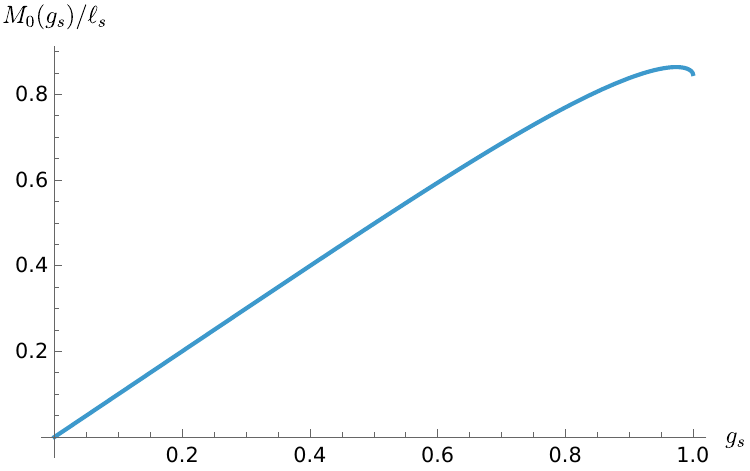}
\caption{The minimum mass as a function of the parameter $g_s$. The maximum occur at $g_s=0.9736$.}\label{fig1}
\end{figure}

The mass is given by
\begin{equation}\label{mass}
M =\frac{r_h}{2}+\frac{g_s^2\ell_s^2}{2r_h} \ {}_2F_1(-1/2,-1/4,3/4,-r_h^4/\ell_s^4),
\end{equation}
and 
\begin{equation}\label{dM}
   dM=\frac{1}{2}\left(1-g_s^2\sqrt{1+\frac{r_h^4}{\ell_s^4}}\right) dr_h
\end{equation}
The interesting point about this is that the mass goes to zero.
From equation (\ref{dM}), we can see that $M'(r_h)\le0$ for $g_s\ge1$ and then we can have $M=0$. For $g_s<1$, we have a minimum mass of the black string at
\begin{equation}
    r_0=\frac{\ell_s g_s}{(1-g_s^4)^{1/4}}.
\end{equation}
The minimum mass exist at $g_s=1$:
\begin{equation}
\lim_{g_s\rightarrow 1}M_0(g_s)=\ell_s\Gamma(3/4)^2/\Gamma(1/2).
\end{equation}
Note that there is no horizon in this case, since $r_0\rightarrow\infty$ where $g_s\rightarrow1$.
Additionally, there exists a value of the parameter $g_s$ such that the minimum mass has a significant value (see Figure (\ref{fig1}). 

We now turn to the study of thermodynamics. Although the Hawking temperature is strictly valid only in the regime \(r_h \gg \ell_s\), its analytical expression supports our interpretation of the parameter \(a\). 

The Hawking temperature is given by
\begin{equation}
T_H=\frac{1}{4\pi}\left(\frac{df}{dr}\right)_{r=r_h}.
\end{equation}
Using the eq. (\ref{frsol}), we obtain
\begin{equation}\label{TH}
    T_H = \frac{1}{4\pi r_h} \left(1 - g_s^2 \sqrt{1 + \frac{\ell_s^4}{r_h^4}}\right).
\end{equation}
For \( g_s^2 \geq 1 \), the Hawking temperature becomes \( T_H \leq 0 \), which is unphysical, while for \( g_s^2 < 1 \), we have \( T_H > 0 \), identifying this as the physically relevant regime. This further supports our interpretation \( |a| = g_s^2 < 1 \), and we will adopt this form from now on. Regarding entropy, we obtain the exact result \( S = \pi r_h^2 \), which is encouraging since this expression has been derived from string theory in various independent frameworks.

For cylindrical symmetry, the metric is:
\begin{equation}\label{cilinder}
    ds^2 = -f(r)dt^2 + \frac{dr^2}{f(r)} + \alpha^2r^2dz^2 + r^2d\theta^2.
\end{equation}
The stress-energy tensor components are:
\begin{align}
    T^{0}_{\ 0} &= -\frac{g_s^2}{r^4}\sqrt{\ell_s^4 + r^4} = T^{1}_{\ 1}, \nonumber \\
    T^{2}_{\ 2} &= \frac{g_s^2\ell_s^4}{r^4\sqrt{\ell_s^4 + r^4}} = T^{3}_{\ 3}. \label{TCY}
\end{align}

The solution for \(f(r)\) is:
\begin{equation}
    f(r) = \frac{2M}{r} + \frac{g_s^2\ell_s^2}{r^2} \ {}_2F_1\left(-\frac{1}{2}, -\frac{1}{4}, \frac{3}{4}, -\frac{r^4}{\ell_s^4}\right),
\end{equation}
and, as expected, black holes are not allowed in cylindrical symmetry.

\section{Conclusions}
As a concluding remark, our results reveal that the extended model preserves key physical properties, such as \(\gamma < 0\), while exhibiting novel features not present in the original Letelier solution. The stress tensor is uniquely determined and differs from the fluid of strings of Letelier in the sense that the pressures are uniquely fixed. For instance, the perfect fluid case, as considered by Soleng in Ref.~\cite{Soleng:1993yr}, is not allowed.

The black hole solution arising from our generalized stress-energy tensor is asymptotically flat and exhibits two horizons. In Letelier’s original work, the parameter \( a \) lacked a dynamical origin and had to be constrained manually to ensure energy positivity and the existence of horizons. In contrast, within our framework, the complete form of the stress-energy tensor naturally enforces the condition \( 0 < a < 1 \). This constraint follows directly from the identification \( a \sim g_s^2 \), consistent with the underlying string-theoretic interpretation. Moreover, a key outcome of our full solution is the realization that Letelier’s model emerges as a particular limit of our construction—specifically, the point-particle approximation of an extended string cloud configuration.

We point out that some caution is required when interpreting the thermodynamic analysis, since Eq.(~\eqref{TH}) suggests the existence of a remnant with nonzero radius, given by
\begin{equation}\label{r0}
    r_h = r_0 = \frac{|\ell_s|}{(\frac{1}{g_s^{4}} - 1)^{1/4}}.
\end{equation}
The presence of a remnant with large entropy could indicate a pathology in the model, potentially excluding it on physical grounds. However, as previously discussed, the Hawking temperature is not valid in this regime, and both backreaction and quantum corrections must be considered before drawing any definitive conclusions. In fact, the inconsistency already appears in Eq.~\eqref{r0} itself, as the resulting remnant radius would lie below the string scale, \( r_0 < \ell_s \), where the semiclassical approximation breaks down.

An interesting extension of this work would be to analyze the rotating case. A rotating counterpart of the present black hole solution would enable the study of black hole shadows, where the influence of the additional hair parameters on the shape and size of the shadow could offer valuable insights into the observational signatures of such exotic geometries. Future investigations could also explore other potential imprints, including gravitational lensing, accretion disk dynamics, gravitational wave emission, and possible connections to dark matter phenomenology. Another promising direction is to examine these solutions within the broader context of modified gravity theories and cosmology.

\section*{Acknowledgments}
\hspace{0.5cm} We sincerely want to thank the referee for their insightful and constructive comments. The critical observations and thoughtful suggestions prompted a thorough reevaluation of our original approach, leading us to undertake a major manuscript revision. The authors thank the Conselho Nacional de Desenvolvimento Cient\'{i}fico e Tecnol\'{o}gico (CNPq) and Fundação Cearense de Apoio ao Desenvolvimento Científico e Tecnológico.

\bibliography{refArxiv}
\end{document}